\documentclass[twoside,twocolumn]{revtex4}
\begin{document}
%\frontmatter

\title{Instantons in the non-linear sigma model}

\author{H.~Albert\vspace{0.2cm}}
%\date{}

\affiliation{\text{II}.Institut f\"ur Theoretische Physik, Universit\"at Hamburg,\\
       Luruper Chaussee 149, 22761 Hamburg, Germany}
%\affiliation{Hamburg University, Germany}

%%%%%%%%%%%%%%%%%%%%%%%%%%%%%%%%%%%%%%%%%%%%%%%%%%%%%%%%%%%%%%
%
% Abstract
%
%%%%%%%%%%%%%%%%%%%%%%%%%%%%%%%%%%%%%%%%%%%%%%%%%%%%%%%%%%%%%%
\begin{abstract}
This is a review of the work done to classify all finite energy solutions of the two dimensional non-linear sigma model. These solutions could be important in understanding the vacuum structure of the non-linear sigma model.
\end{abstract}
\maketitle
%%%%%%%%%%%%%%%%%%%%%%%%%%%%%%%%%%%%%%%%%%%%%%%%%%%%%%%%%%%%%%
\section{Introduction}
%%%%%%%%%%%%%%%%%%%%%%%%%%%%%%%%%%%%%%%%%%%%%%%%%%%%%%%%%%%%%%
The non-linear sigma model in two dimensions [1,2] is a field theoretical description of the Heisenberg ferromagnet. For sufficiently low temperature the interaction between the local magnets become dominant and the local magnets tend 
to align. Thus we get an ordered state characterized by an order parameter, which can be chosen to be the direction of the local spin vector. Thus the order parameter is a unit vector, which can be represented by a triple of scalar fields $[\varphi^1 (x) ,\varphi^2 (x) ,\varphi^3 (x)]$, subject to the constraint $\varphi^b \varphi^b = 1 $. 
The static energy functional of the Heisenberg ferromagnet is given by 
$$H[\varphi ] = \langle d\varphi^b | d\varphi^b \rangle = \int \partial_j \varphi^b \partial^j \varphi^b \sqrt{g} d^D x$$ 
with 
$$\varphi^b \varphi^b = 1\text{.}$$ 
Since we want to classify all finite energy solutions of the non-linear sigma model, we have to impose $$H[\varphi ] = \langle d\varphi^b | d\varphi^b \rangle  < \infty$$which leads to the boundary condition$$\lim_{x\to\infty} d\varphi^b (x) = 0$$ 
and that $\varphi^b$ must be asymptotically constant [11] 
$$\lim_{x\to\infty} \varphi^b (x) = \varphi_0^b\text{.}$$ 
Which components the constant vector at infinity has cannot be said. 
All unit vectors are equally likely. 
Hence, we follow convention and choose the north pole as the asymptotic value
$$\lim_{x\to\infty} \varphi^b = [0;0;1]\text{.}$$ 
From the very beginning we chose the vector $\varphi^b $ to be a triple. But nothing prevents us from  asking whether we can extend the model with $\varphi^b $ being a N-tuple. But in the next section we will see with the help of homotopy theory that for N $>$ 3 there are no nontrivial vacuum solutions. Finally it should be said that the global (non space-time dependent) symmetry group of the non-linear sigma model is O(N) for $\varphi^b$ being a N-tuple or especially O(3) 
for the Heisenberg ferromagnet, i.e. $\varphi$ being a triple. 
As a final comment we should say that we will not discuss generalized non-linear sigma models, 
where space-time and target (field)-space may have arbitrary geometries [7].
%
%
%%%%%%%%%%%%%%%%%%%%%%%%%%%%%%%%%%%%%%%%%%%%%%%%%%%%%%%%%%%%%%
\section{Instantons and the non-linear sigma model }
%%%%%%%%%%%%%%%%%%%%%%%%%%%%%%%%%%%%%%%%%%%%%%%%%%%%%%%%%%%%%%
In this section I will make use of the exterior derivative calculus, since use is made of
the duality operation, which can be written best in these terms.
Due to the constraint $\varphi^2= 1$, the scalar field in the non-linear sigma model
(or Heisenberg ferromagnet) defines a mapping 
$$\varphi^a :R^2 \rightarrow S^{N-1}\text{.}$$
We can compactify $R^2$ by making use of the boundary 
condition (see above in the introduction) on $\varphi^a$:
$$\lim_{x\to\infty} \varphi^a (x) = (1, 0, 0..., 0)$$
All points at infinity on $R^2$ can be identified, giving a map $f: R^2 \rightarrow S^2$
lifting the map $\varphi^a$ to the map $\varphi^a: S^2 \rightarrow S^{N-1}$.
But that means, that $\varphi^a$ is a representative of an element of $\pi_2 (S^{N-1} )$,
where $\pi_2 (S^{N-1} ) = 0$ for N equal or larger than 4.
This can best be seen by using fiber bundle techniques:
Define a principal fiber bundle: 
$$O (N) \stackrel{\pi}{\longrightarrow} \frac {O (N)}{O (N-1)}\text{,}$$ with fiber O (N-1). Now
O (N) is the translational group on the sphere $S^{N-1}$, while O (N-1) leaves a point on the sphere fixed.
Hence, the quotient space $O (N)/O (N-1)$ is isomorphic to $S^{N-1}$.
We then have an exact sequence of homotopy groups:
$$...\longrightarrow \pi_i (O (N-1)) \stackrel {\alpha}{\longrightarrow} \pi_i (O(N) ) \longrightarrow \pi_i (S^{N-1} )$$
                                                      $$  \longrightarrow \pi_{i-1} (O(N-1)) \stackrel {\beta}{\longrightarrow} ...$$
where $\alpha$ is an epimorphism (surjective), $\beta$ is an isomorphism for $i \le N-2$.                                                         
Since the sequence is exact, it follows, that $\pi_i (S^{N-1}) = 0$ for $i \le N-2$ (see [5,6]).
Restricting  $\varphi$ to be a triple, and hence the symmetry group of the sigma model being O(3), it is a representative of an element of $\pi_2 (S^2)$ ($\pi_2$ else being zero, as explained above), the space of solutions
is divided into sectors, labeled by n, meaning the number, specifying, how often $\varphi$ maps $S^2$ on $S^2$, called the mapping
degree or winding number.
The Hamiltonian of the $\sigma$- model can be cast into a form, exhibiting directly the splitting of
the space of solutions into sectors, labeled by n:
$$H[\varphi] = \frac{1}{4} ||*d\varphi^a \mp \# d\varphi^a ||^2 \pm 4\pi
n(\varphi )\text{,}$$ where the $*$ operation means the Hodge dual of the one form and the
operation $\#$ is defined in the section ``Duality condition''.
The derivation of the Hamiltonian runs as follows: For it we need two more concepts:\begin{itemize} \item{The selfduality equation, including conformal invariance} and \item{an analytical expression for the winding number or, what is the same, the topological charge.}\end{itemize}
\section{Duality condition}
The derivatives $\partial_i \varphi$ are tangent vectors to the sphere $S^2$ in field space, made
up by the fields $\varphi^a$ due to the constraint $\varphi^2 = 1$. We introduce a duality
operation, which will be called $\#$, by the formula 
$$\# d\varphi^a =\epsilon_{abc} \varphi^b d\varphi^c\text{.}$$
This corresponds to a rotation by $\frac{\pi}{2}$ in the tangent space to the sphere $S^2$ in field
space, described above.
%%%%%%%%%%%%%%%%%%%%%%%%%%%%%%%%%%%%%%%%%%%%%%%
\section{winding number}
The winding number tells us, how often a topological space (not necessarily $S^N$) is mapped on another
topological space. One example is the Cauchy formula. The crucial point is that the mapping from one topological space to another must not be continuously be deformed to the 
constant mapping (singular transformations alter the winding number by changing the
topology of the target space). If this is possible, the winding number is zero.
%%%%%%%%%%%%%%%%%%%%%%%%%%%%%%%%%%%%%%%%%%%%%%%
\section{The winding number or The Brouwer theorem}
To introduce the Brouwer index, we need the concept of the degree of a map (smooth):
\\{\bf Definition}:
\\The degree of a smooth map $f: \cal M \longrightarrow \cal N$, $\cal M , \cal N$ being compact, orientable manifolds of the same dimension, at a regular value $Q \epsilon \cal N$
 is the integer $$Deg (f;Q) = \sum_{P_i \epsilon f^{-1} (Q)} sgn | \partial y^i /\partial x^j |_{P_i}\text{,}$$
 where $sgn |\partial y^i /\partial x^j | = \pm  1$ according to whether $f^*$ preserves or reverses orientation.
 \\{\bf Brouwer's theorem}:
\\ Let $f: \cal M \longrightarrow \cal N$ be be a smooth map, T a n-form on $\cal N$. Then:
 $$\int_{\cal M} f^* T = Deg (f) \int_{\cal N} T\text{.}$$
 Now lets choose T as $\epsilon = (g)^{1/2} dx^1\wedge ... \wedge dx^n$, the volume form on $\cal N$. Then we get
 $$\int_{\cal M} f^* \epsilon = Deg(f)\int_{\cal N} \epsilon = Deg(f) Vol[\cal N]$$ or
 $$Deg(f) = \frac{\int_{\cal N} f^* \epsilon}{Vol[\cal N]}$$
 The map in our $\sigma$ model is given by: $\varphi^a : R^2 \longrightarrow S^{N-1}$, $\varphi^2 = 1$
 and the winding number by 
 $$Deg(\varphi) = \frac{1}{Vol(S^{N-1})} \int_{R^2} \epsilon_{a_0...a_n} 
 \varphi^{a_0} d\varphi^{a_1} ...d\varphi^{a_n}\text{.}$$
 We know now, that $\pi_2 (O(N))$ is zero for N larger than 3, so we restrict to the case N = 3.
 $$\Rightarrow Deg(\varphi ) = n = \frac{1}{8\pi} \int_{S^2} \epsilon_{abc} \varphi^a d\varphi^b \wedge
 d\varphi^c\text{,}$$ 
 in coordinates $n = \frac{1}{4\pi} \int \epsilon_{abc} \epsilon_{ij} \varphi^a \partial_i \varphi^b \partial_j \varphi^c dx^1 dx^2 $,
 or 
 $$n = \frac{1}{8\pi} <*d\varphi^a |\# d\varphi^a>\text{.}$$
 Finally, we observe $$<d\varphi^a | d\varphi^a> = <*d\varphi^a |*d\varphi^a> = <\# d\varphi | \#
 d\varphi^a>\text{,}$$
 since as already said above a
 duality transformation in two dimensions amounts to a rotation about $\frac{\pi}{2}$, 
 which leaves the norm ($||\;\;||^2 =\langle | \rangle$)
 invariant.
 Now we have all the ingredients, to invoke the Bogomolny decomposition.
 $$H = \frac{1}{2} <d\varphi^a | d\varphi^a>$$ $$= \frac{1}{4} <*d\varphi^a | *d\varphi^a > + \frac{1}{4} <\# d\varphi^a |\# d\varphi^a> +$$ $$ \frac{1}{2} <*d\varphi^a |\# d\varphi^a> - \frac{1}{2} <*d\varphi^a | \# d\varphi^a>$$
$$= \frac{1}{2} <*d\varphi^a - \# d\varphi^a |*d\varphi^a - \# d\varphi^a> + 4\pi n(\varphi)$$ 
$$= \frac{1}{2} ||*d\varphi^a - \# d\varphi^a ||^2 + 4\pi n(\varphi)$$
This form of the Hamiltonian is called the Bogomolny decomposition.
From this formula follows\begin{itemize}
\item{ The energy is bounded below by $4\pi n(\varphi)$}
\item {A configuration with winding number n is a ground state if and only if it satisfies the first order
   differential equation $$*d\varphi^a = \# d\varphi^a$$ if n is positive and
   $$*d\varphi = -\# d\varphi^a$$ if n is negative.}\end{itemize} 
   Ground state solutions in the sigma model are called spin waves.
   Pictorially the winding number (or topological charge) describes, how often the spins, aligned along the, say, x-axis, twist
   around this axis, the whole chain of spins being held fixed at both "ends" by boundary conditions.
   These equations are so called double self-dual equations and play a decisive role in classifying the solutions.
The next step will be the proof of the following\\ {\bf Proposition}:
\\A spin configuration $\varphi^a : R^2 \longrightarrow S^2$ is a spin wave if and only if it is a conformal
map from the plane to the sphere.
\\Proof:
\\Let $e_1, e_2$ be a canonical basis of $R^2$. These are lifted to the tangent vectors $\partial_1 \varphi^a , \partial_2 \varphi^a$.
If $\varphi^a$ is a conformal map, this forces $\partial_1 \varphi^a , \partial_2 \varphi^a$ to be
orthogonal and of the same length.
Now we know, that a solution $\varphi^a$ with winding number n is a solution of the following
equations: $$\partial_1 \varphi = \# \partial_2 \varphi , \partial_2 \varphi = -\# \partial_1 \varphi$$
But this shows, that the vectors $\partial_i \varphi^a$ are orthogonal, since the operator
$\#$ amounts to a rotation of $\frac{\pi}{2}$, as already said above. They are also of the same length, since
$\#$ does not alter the length. We now have to show the converse, i.e., that any conformal map
solves the two differential equations above:
Let $\varphi^a : R^2 \longrightarrow S^2$ be a smooth map and introduce the following vectors:
$$P^a = *d\varphi^a - \# d\varphi^a  i.e. P_i = \epsilon_{ij} \partial_j \varphi - \varphi \times \partial_i \varphi$$
$$Q^a = *d\varphi^a + \# d\varphi^a i.e. Q_i = \epsilon_{ij} \partial_j \varphi + \varphi \times \partial_i \varphi$$
Then we get $$P_1 Q_1 = P_2 Q_2 = |\partial_1 \varphi |^2 - |\partial_2 \varphi |^2$$ $$P_1 P_2 = Q_1 Q_2 = 0$$
and 
$$P_1 Q_2 = P_2 Q_1  = -2\partial_1 \varphi \partial_2 \varphi\text{.}$$ 
But for a conformal map $\varphi : R^2 \longrightarrow S^2$, $\partial_1 \varphi , \partial_2 \varphi $ are orthogonal tangent vectors of the same length and hence the right hand sides of the equations above vanish. This shows that the $P_1 , P_2 , Q_1 , Q_2$ are mutually orthogonal.
The next and final step is to show that\\{\bf Proposition}\\ If $\varphi$ is a conformal map the either 
$$P_1 = P_2 = 0$$
or
$$Q_1 = Q_2 = 0\text{.}$$
For this we need the following {\bf remark}:
If $\partial_1 \varphi$ (or $\partial_2 \varphi $) vanishes, then all four vectors $P_1, P_2, Q_1, Q_2$ vanish. To
see this, let $\partial_1 \varphi$ vanish. Then we get $$P_2 = - \varphi \times \partial_2 \varphi$$and
         $$Q_2 = \varphi \times \partial_2 \varphi$$ But since they are orthogonal vectors, $\partial_2 \varphi$ has also to vanish.
Clearly $P_1, P_2, Q_1, Q_2$ all have to vanish.
To prove the proposition we observe, that $P_1, P_2, Q_1, Q_2$ all are tangent vectors in the same two-dimensional
plane. Since they are all mutually orthogonal, two of them have to vanish. But we have to exclude the possibilities, that either
two $P_i, Q_j$ vanish, leaving the other two nonzero.
Suppose now, that $P_1, Q_1$ is equal to zero. Then $\partial_2 \varphi =\frac{1}{2} (P_1 + Q_1 )$ vanishes and hence all $P_i , Q_j $ vanish. The same amounts for $P_2 , Q_2 $.  Suppose now that $Q_2 = P_1 = 0$. From $Q_2 = 0$ we get $\partial_1 \varphi = \varphi \times \partial_2 \varphi$ .
Inserting this into the expression for $P_1$, we obtain
$$0 = P_1 = \partial_2 \varphi - \varphi \times (\varphi \times \partial_2 \varphi) = 2\partial_2 \varphi\text{.}$$
So $2\partial_2 \varphi$ vanishes as well and so $P_1, P_2, Q_1, Q_2$. The same is for $Q_1 = P_2 = 0$
cases and we are left with $$P_1 = P_2 = 0$$ or $$Q_1 = Q_2 = 0$$ for
$\varphi : R^2 \longrightarrow S^2$ being conformal. This proves the proposition.
In the final step, we introduce complex analysis. Orientation preserving conformal maps $\varphi^b :S^2 \longrightarrow S^2$
 are necessarily algebraic, that means: $$\varphi = \frac{P(z)}{Q(z)}\text{,}$$ with
P, Q  arbitrary polynomials [8].
These correspond to solutions with positive winding number. Negative winding number
solutions are represented by antiholomorphic (orientation reversing) maps.
The winding number is given by the degree of the polynomial P(z). This can be explained as follows:
Let $w_0$ be a regular value, i.e. $| \frac{\partial_i y}{\partial_j x} | \neq 0$. It follows, that
$$P(z) - w_0 Q(z) = 0$$ has n different solutions (n being the degree of P(z)). But this is by 
definition the winding number.
Finally, we have to clarify, whether there could be other finite energy solutions than the spin waves.
The answer is negative as shown by G. Woo [2].
As a final remark, lets have a look at the stereographic projection $\pi$
$$w = \frac{\varphi^1}{1 -\varphi^3} + i\frac{\varphi^2}{1 -\varphi^3}$$ from two dimensional
sphere down to the complex plane. In the same way the tangent vectors $\partial_i \varphi^a$ are
on the unit sphere are projected down into $\partial_i w$ in the complex plane, i.e.,
$$\pi^* : d\varphi^a \longrightarrow dw\text{.}$$
Since $\pi$ is conformal, it preserves right angles, but it reverses orientation:
$$\epsilon_{abc} \varphi^b \partial_i \varphi^c \longrightarrow - i \partial_i w\text{,}$$ i.e.,
$$ \# d\varphi^a \longrightarrow - i dw\text{.}$$
From linearity, it follows: 
$$\epsilon_{ij} \partial_j  \varphi^a \longrightarrow \epsilon_{ij} \partial_j w\text{,}$$ i.e.,
$$*d\varphi^a \longrightarrow *dw\text{.}$$
Putting all this together, we see that the double self duality equation is projected
down to the self dual equation $$*dw =-idw\text{.}$$ 
This equation is conformal invariant in two
dimensions (and 4). Complex analysis tells us, that any solution $dw$ gives automatically a
holomorphic function $w$ and vice verse.
There seems to be a contradiction to Derrick's scaling argument [9] in that we have stable finite
energy solutions. But this is only apparent, since the sigma model represents a loophole
in Derrick's argumentation in not having a potential term in the Lagrangian. Generally pure
scalar field theories have only soliton solutions in dim = 1.
Derrick's argumentation runs as follows:
Consider a pure scalar field theory in D dimensions with Lagrange density
$$L = -1/2 \partial_{\mu} \varphi^a \partial^{\mu} \varphi^a - U(\varphi^a)\text{.}$$
The corresponding static energy functional is given by
$$H(\varphi^a ) = 1/2 \int_{R^d} \partial_i \varphi^a \partial^i \varphi^a + \int_{R^D} U(\varphi^a )$$
$$= H_1 (\varphi ) + H_2 (\varphi)\text{.}$$
Suppose $\varphi^a (x)$ is a static solution and consider the scaled configuration
$$\varphi^a_{\lambda} (x) = \varphi^a (\lambda x)\text{.}$$
The scaled configuration has static energy
$$H(\varphi^a_{\lambda} = 1/2 \int \lambda^2 \partial_i \varphi^a (\lambda x) \partial^i \varphi^a (\lambda x)d^D x$$ $$ + \int U(\varphi^a (\lambda x) d^D x$$
$$= 1/2 \int \lambda^{2-D} \partial_i \varphi^a (y) \partial^i \varphi^a (y) d^D y $$ $$+ \lambda^{-D} \int U(\varphi^a (y)) d^D y$$
$$= \lambda^{2-D} H_1 + \lambda^{-D} H_2\text{.}$$
If $\varphi^a$ is to be stable, $H$ must be stationary against variations of $\lambda$:
$$0 = \partial_{\lambda} H_{\lambda = 1} = (2 - D)H_1 - DH_2$$
This implies: A nontrivial static solution in a pure scalar field theory is unstable, if
the space dimension exceeds 2.
For dimension two we have $H_2 (\varphi^a ) = 0$, which can be fulfilled, if the potential
energy term is simply absent. This is the case in the two dimensional non-linear sigma model.
%%%%%%%%%%%%%%%%%%%%%%%%%%%%%%%%%%%%%%%%%%%%%%%%%%%%%%%%%%%%%%
\section{Vacuum structure}
A field theory which satisfies the conditions of Lorentz (Euclidean) invariance, 
spectrum and locality, the vacuum is unique if and only if the n-point functions have 
the cluster decomposition property [10]. Y.~Iwasaki shows [4] that the correlation 
functions have the cluster property for instanton or anti-instanton contributions but not 
for instanton - anti-instanton contributions. 
Similar ideas concerning the cluster property and topological non-trivial solutions [12], 
instantons, apply to Yang-Mills theory [13].

%%%%%%%%%%%%%%%%%%%%%%%%%%%%%%%%%%%%%%%%%%%%%%%%%%%%%%%%%%%%%%
%
% Bibliography
%
%%%%%%%%%%%%%%%%%%%%%%%%%%%%%%%%%%%%%%%%%%%%%%%%%%%%%%%%%%%%%%

\end{document}